\documentclass[nofootinbib,prl,aps,reprint,amsmath,amssymb]{revtex4-1}
\usepackage{hyperref}
\usepackage{changes}
\usepackage[section]{placeins}
\usepackage[titletoc]{appendix}
\usepackage{xcolor}
\usepackage{dsfont}
\usepackage{bm}
\usepackage{mathtools}
\usepackage{enumerate}
\DeclareMathAlphabet{\mathpzc}{OT1}{pzc}{m}{it}

\newcommand{\sayy}[1]{`#1'}
\providecommand{\href}[2]{#2}

\def\be{\begin{equation}}
\def\ee{\end{equation}}
\def\bea{\begin{eqnarray}}
\def\eea{\end{eqnarray}}

\def\sig{\sigma}

\def\la{\langle}
\def\ra{\rangle}
\def\Eu{ \mathfrak{H} }
\def\Sp{ \mathcal{S} }
\def\obs{\mathcal{O}}
\def\emi{\mathcal{E}}
\def\lie{\mathcal{L}}
\definecolor{MyB}{rgb}{0.1,0.1,1.0}

\begin{document}
\title{Multipole decomposition of redshift drift -- model independent mapping of the expansion history of the Universe} 
\author{Asta~Heinesen}
\email{asta.heinesen@ens--lyon.fr}
\affiliation{Univ Lyon, Ens de Lyon, Univ Lyon1, CNRS, Centre de Recherche Astrophysique de Lyon UMR5574, F--69007, Lyon, France}

\begin{abstract}
We consider redshift drift in a general space-time as expressed in terms of physically interpretable multipole series. 
An important realisation from the derived results is that redshift drift cannot in general be thought of as a direct probe of the average expansion rate of the Universe due to the presence of structure along the light beams from the astrophysical sources to the observer. We also predict the general presence of a dipolar and a quadrupolar offset in the detection of redshift drift for observers placed in locally anisotropic environments. 
\end{abstract}

\maketitle

\section{Introduction}
%%%%%%%%%%%%%%%%%%%%%%%%%%%%%%%%%%%%%%%%%%%%%%%%%%%%%%%%%%%%%%%%%%% 
Redshift drift -- denoting the change of redshift of light from an astrophysical source in time -- is a promising probe of the cosmological expansion history. 
Detection of redshift drift allows for directly inferring kinematic properties of the Universe, which would otherwise require model assumptions. The observation time required to detect the signal of redshift drift has been estimated to be one to a few decades depending on the instrumentation used \cite{Balbi:2007fx,Liske:2008ph}. 
Redshift drift has been studied for comoving observers within the Friedmann-Lema\^{\i}tre-Robertson-Walker (FLRW) class of metrics, see e.g. \cite{Sandage,McVittie,Loeb:1998bu}, and the effect of peculiar motion relative to the FLRW \sayy{background} model has been investigated \cite{Killedar:2010,Bolejko:2019vni}. 
Predictions for redshift drift have been formulated within spherically symmetric space-times \cite{Uzan:2008qp} and in Lema\^{\i}tre-Tolman-Bondi models \cite{LTB_zdrift} with the observer placed in the center of symmetry. Redshift drift has also been studied within classes of Szekeres cosmological models \cite{Mishra:2014vga,Mishra:2012vi,Mishra:2013iu}, in Stephani models \cite{Balcerzak:2012bv}, and in Bianchi I models \cite{Fleury:2014rea}. 
The drift of redshift has been considered for space-time congruences of vanishing acceleration \cite{Koksbang:2015ctu,Jimenez:2019cll}, and its conjectured average applicable to a set of many observers and emitters \cite{Koksbang:2015ctu} has been tested for certain numerical space-time solutions satisfying a notion of statistical homogeneity and isotropy \cite{Koksbang:2020zej,Koksbang:2019glb}. 
Redshift drift and its relation to position drift has been considered in general space-times \cite{Korzynski:2017nas}.  

In this paper we propose a formalism which allows for analysing the impact of local anisotropies and inhomogeneities in the Universe on the measured drift of redshift without imposing model assumptions.  
We show that regional anisotropies along the null rays in general contribute systematically to the measured redshift drift, also for space-times with a notion of large scale homogeneity and isotropy. As a consequence, redshift drift cannot in general be thought of as a direct probe of the average expansion rate of the Universe. 
The results presented are relevant for analysis strategies that aim at model independent determination of the expansion history of the Universe through measurements of redshift drift.

\vspace{5pt} 
\noindent
\underbar{Notation and conventions:}
Units are used in which $c=1$. Greek letters $\mu, \nu, \ldots$ label spacetime
indices in a general basis. Einstein notation is used such that repeated indices are summed over.  
The signature of the spacetime metric $g_{\mu \nu}$ is $(- + + +)$ and the connection $\nabla_\mu$ is the Levi-Civita connection. 
Round brackets $(\, )$ containing indices denote symmetrisation in the involved indices and square brackets $[\, ]$ denote anti-symmetrisation. 
Bold notation $\bm V$ for the basis-free representation of vectors $V^\mu$ is used occasionally. 
%%%%%%%%%%%%%%%%%%%%%%%%%%%%%%%%%%%%%%%%%%%%%%%%%%%%%%%%%%%%%%%%%%%%
%%%%%%%%%%%%%%%%%%%%%%%%%%%%%%%%%%%%%%%%%%%%%%%%%%%%%%%%%%%%%%%%%%%  

\section{Redshift drift in a general space-time}
\label{sec:zdrift}
%%%%%%%%%%%%%%%%%%%%%%%%%%%%%%%%%%%%%%%%%%%%%%%%%%%%%%%%%%%%%%%%%%% 
We consider an arbitrary space-time with an observer and an emitter with worldlines $\gamma_o$ and $\gamma_e$ generated by four velocity vectors $\bm{u}_o$ and $\bm{u}_e$ respectively, and with proper times $\tau_o$ and $\tau_e$ parametrising the worldlines. 
Let $\bm k$ be the 4--momentum of a non-caustic geodesic null congruence with null lines generating a bijective map between the observer and the emitter worldlines, defining the invertible function $\tau_{o} \mapsto \tau_e(\tau_{o})$ between the proper times of the worldlines. We define the differential operator $\frac{d}{d\tau_o}$ through its acting on functions $\tau_{o} \mapsto \mathcal{F}(\tau_{o})$ and $\tau_{e} \mapsto \mathcal{G}(\tau_{e})$ defined on $\gamma_o$ and $\gamma_e$ respectively: 
\bea
\label{derivative}
\hspace*{-0.2cm} \frac{d \mathcal{F} (\tau_{o}) }{d\tau_o}  \equiv    u_o^\mu \nabla_\mu \mathcal{F}(\tau_{o})   \, , \quad \frac{d \mathcal{G} (\tau_{e}) }{d\tau_o}  \equiv   \frac{d \tau_e }{d  \tau_o} u_e^\mu \nabla_\mu \mathcal{G}(\tau_{e})     \, , 
\eea 
where $\tau_{o} \mapsto \mathcal{G}(\tau_{e}(\tau_o))$ is the composite function defined through the bijective map of the null congruence. 
The redshift function of null rays propagating from the emitter to the observer is given by 
\bea
\label{redshift}
&& z(\tau_o) \equiv  \left( \frac{d \tau_e (\tau_o) }{d  \tau_o}  \right)^{\! -1}   - 1 \equiv  \frac{E(\tau_e (\tau_o))}{E_o (\tau_o)}  - 1 \, ,
\eea 
where the energy functions of the emitted and observed photons 
\bea
\label{energyfun}
&& E_e(\tau_e) \equiv -  u_e^\mu k_\mu (\tau_e)  \, , \qquad   E_o(\tau_o) \equiv -  u_o^\mu k_\mu (\tau_o) \, 
\eea 
are defined at $\gamma_e$ and $\gamma_o$ respectively. 
The Jacobian $\frac{d\tau_e}{d\tau_o}$ has interpretation as the ratio of the photon wavelengths measured by the emitter and observer.  
We now consider a point of observation of a null ray $\obs$ intersecting $\gamma_o$ -- with the same null ray intersecting $\gamma_e$ at the point $\emi$ -- and define redshift drift as the change in redshift induced by a small increase in proper time of the observer  
\bea
\label{redshiftdrift}
\hspace*{-0.3cm} \frac{d z}{d \tau_o} \Bigr\rvert_{\obs} &=&    - (1 + z) \frac{ u_o^\mu \nabla_\mu E_o (\tau_o) }{E_o (\tau_o)}  \Bigr\rvert_{\obs}  +  \frac{ u_e^\mu \nabla_\mu E_e (\tau_e) }{E_e (\tau_e)}  \Bigr\rvert_{\emi}  \, , 
\eea 
where we have used the definitions (\ref{derivative}) and (\ref{redshift}). 

\section{Multipole decomposition of the redshift drift}
\label{sec:multipole} 
Let us consider the case where the observer and emitter worldlines are embedded in a space-time congruence of worldlines generated by the 4-velocity field $\bm u$. 
We can can rewrite the redshift drift function (\ref{redshiftdrift}) by use of the generic decomposition 
\bea
\label{kdecomp}
k^\mu = E(u^\mu - e^\mu) \, , \qquad  e^\mu e_\mu = 1 \, , 
\eea  
where $-\bm e$ is the spatial propagation direction of the null ray relative to the observer comoving with $\bm u$. 
This yields 
\bea
\label{redshiftdriftdec}
\frac{d z}{d \tau_o} \Bigr\rvert_{\obs} = (1+z) \Eu_\obs  - \Eu_\emi + \Sp_{\emi \rightarrow \obs}  \, ,
\eea  
where
\bea
\label{def:Eevolution}
 \Eu \equiv - \frac{ k^{\mu}\nabla_{\mu} E }{E^2} =  \frac{1}{3}\theta  - e^\mu a_\mu + e^\mu e^\nu \sigma_{\mu \nu}   \, 
\eea  
is the rate of change of the photon energy along the null lines, and where\footnote{The term $\Sp_{\emi \rightarrow \obs}$ does not appear in the expressions for redshift drift in \cite{Koksbang:2015ctu,Jimenez:2019cll}, where the derivations implicitly assume contributions from the spatial derivative of the energy function $e^\mu \nabla_\mu E$ to vanish. As we shall discuss in the following, the term $\Sp_{\emi \rightarrow \obs}$ contains contributions that cannot be assumed to be subdominant in the general analysis of redshift drift measurements.} 
\bea
\label{def:Spatialder}
\Sp_{\emi \rightarrow \obs} \equiv -  (1+z) \frac{e^\mu \nabla_\mu E}{E}  \Bigr\rvert_{\obs}  + \frac{e^\mu \nabla_\mu E}{E}  \Bigr\rvert_{\emi}   \, 
\eea  
is the contribution from the change of the energy function along the spatial direction of propagation of the photon. 
In deriving (\ref{def:Eevolution}) we have used the definition of the energy function $E\equiv - \bm u \cdot \bm k$ along with the general decomposition  
\bea
\label{def:expu}
&& \nabla_{\nu}u_\mu  = \frac{1}{3}\theta h_{\mu \nu }+\sig_{\mu \nu} + \omega_{\mu \nu} - u_\nu a_\mu  \ , \nonumber \\ 
&& \theta \equiv \nabla_{\mu}u^{\mu} \, ,  \quad \sig_{\mu \nu} \equiv h_{ \la \nu  }^{\, \beta}  h_{  \mu \ra }^{\, \alpha } \nabla_{ \beta }u_{\alpha  }  \, , \nonumber \\ 
&& \omega_{\mu \nu} \equiv h_{  \nu  }^{\, \beta}  h_{  \mu }^{\, \alpha }\nabla_{  [ \beta}u_{\alpha ] }   \, , \quad  a^\mu \equiv u^\nu \nabla_\nu u^\mu \,  , 
\eea 
where $h_{ \mu }^{\; \nu } \equiv u_{ \mu } u^{\nu } + g_{ \mu }^{\; \nu } $ is the spatial projection tensor defined in the frame of the 4--velocity field $\bm u$ and where triangular brackets $\la \ra$ denote traceless symmetrisation in the involved indices of a tensor in three dimensions\footnote{For two indices we have that the traceless parts of symmetric spatial tensors $T_{\mu \nu} = T_{(\mu  \nu)} = h_{ \mu }^{\, \alpha } h_{ \nu }^{\, \beta } T_{(\alpha  \beta)}$ is given by $T_{\la \mu \nu \ra}  = T_{\mu \nu} - \frac{1}{3}h_{\mu \nu} T^{\alpha}_{\; \alpha}$. Analogously for a tensor with three indices satisfying $T_{\mu \nu \rho} = T_{(\mu  \nu \rho)} = h_{ \mu }^{\, \alpha } h_{ \nu }^{\, \beta } h_{ \rho }^{\, \gamma } T_{(\alpha  \beta \gamma)}$, we have $T_{\la \mu \nu \rho \ra} = T_{\mu \nu \rho} - \frac{1}{5} \left( T_{\mu} h_{\nu \rho}  + T_{\nu} h_{\rho \mu} + T_{\rho} h_{\mu \nu} \right)$, with $T_\mu \equiv T^\nu_{\; \, \nu \mu}$ .}. 
The kinematic variables $\theta$, $\sigma_{\mu \nu}$, and $\omega_{\mu \nu}$ represent the expansion, shear, and vorticity respectively of the observer congruence, while $a^\mu$ is the four-acceleration of the flow-lines of the congruence.  
We note that the rate of change of photon energy (\ref{def:Eevolution}) has interpretation as a multipole series in $\bm e$ that is truncated at the quadrupole level, with $\{\theta/3, -a_\mu , \sigma_{\mu \nu} \}$ representing the monopole, dipole, and quadrupole coefficients of the multipole series respectively. 
The first two terms of (\ref{redshiftdriftdec}) are similar in form to the FLRW expression for redshift drift if $\Eu$ is interpreted as an effective (anisotropic) Hubble parameter. 
The last term of (\ref{redshiftdriftdec}) can be rewritten as the integral expression 
\bea
\label{ederivative}
\hspace*{-0.3cm} \Sp_{\emi \rightarrow \obs}  =   E_\emi \! \! \int_{\lambda_\emi}^{\lambda_\obs} \! \! \! d \lambda \, \mathcal{I} \, , \qquad \mathcal{I} &\equiv& - k^\nu \nabla_\nu \! \! \left( \! \frac{e^\mu \nabla_\mu E}{E^2}  \! \right)  \, , 
\eea  
where $\lambda$ is an affine parameter along the photon geodesic satisfying $k^\mu \nabla_\mu \lambda =1$. 
The integrand in (\ref{ederivative}) reads 
\bea
\label{integrand}
\hspace*{-0.3cm} \mathcal{I}  &=& e^\mu \nabla_\mu \Eu(\bm e) + e^\mu a_\mu \Eu(\bm e)   - \frac{1}{E^2}p^{\mu}_{\; \nu} \lie_{\bm{k}}(e^\nu) \nabla_{\mu}E  \,    , 
\eea  
where $p^{\mu}_{\; \nu} \equiv u^\mu u_\nu - e^\mu e_\nu + g^{\mu}_{\; \nu}$ is the projector onto the lightfront defined by $\bm k$ and $\bm u$. The operator $\lie_{\bm{k}}$ denotes the lie derivative along $\bm k$, and we have used the rewriting  
\bea
\label{lie}
\lie_{\bm{k}}(e^\mu) = p^{\mu}_{\; \nu} \lie_{\bm{k}}(e^\nu)  +     k^\mu e^\nu a_\nu + 2 e^{[ \mu} u^{\nu ]} \nabla_{\nu} E \, 
\eea  
of the lie derivative of $\bm{e}$ along $\bm{k}$, which follows from use of the decomposition (\ref{kdecomp}) and the definition of $p^{\mu}_{\; \nu}$. 
We can write the integrand (\ref{integrand}) as an expansion in the spatial unit vector $e^\mu$ and the projection of its acceleration $d^\mu \equiv p^{\mu}_{\; \nu} e^\alpha \nabla_\alpha e^\nu$ in the following way\footnote{The last term of (\ref{integrand}) can be rewritten using $\frac{1}{E^2} p^{\mu}_{\; \nu}  \lie_{\bm{k}}(e^\nu) = - 2 \frac{1}{E}  p^{\mu}_{\; \nu} e^\alpha \left( \sigma^{\nu}_{\; \alpha} + \omega^{\nu}_{\; \alpha}   \right) + \frac{1}{E}  p^{\mu}_{\; \nu} a^\nu  + \frac{1}{E} p^{\mu}_{\; \nu}  e^\alpha \nabla_\alpha e^\nu$
together with the identity $p^{\mu  \nu} \nabla_\nu E = E p^{\mu}_{\; \nu}  e^\alpha \nabla_\alpha e^\nu - 2 E p^{\mu}_{\; \nu}  e^\alpha \omega^{\nu}_{\; \alpha} +  2 p^{\mu  \nu} u^\alpha \nabla_{[\alpha} k_{\nu]}$ .  } 
\bea
\label{integrandexp}
\hspace*{-0.65cm} \mathcal{I} \! = \! \mathcal{I}^{\it{o}} \! \!+\! e^\mu \mathcal{I}^{\bm{e}}_\mu   \!+\!   d^\mu \mathcal{I}^{\bm{d}}_\mu   \!+\!   e^\mu \! e^\nu \mathcal{I}^{\bm{ee}}_{\mu \nu}  \!+\!  e^\mu \! d^\nu \mathcal{I}^{\bm{ed}}_{\mu \nu}  \!+\!   e^\mu \! e^\nu \!e^\rho \mathcal{I}^{\bm{eee}}_{\mu \nu \rho} 
\eea  
with the non-vanishing multipole coefficients defined as 
\bea
\label{coef}
&& \mathcal{I}^{\it{o}} \equiv - \frac{1}{3} (4 \omega^{\mu \nu} \omega_{\mu \nu}  + D_{\mu} a^{\mu} + a^{\mu} a_{\mu}  )  - d^\mu d^\nu h_{\mu \nu}  \, , \nonumber  \\   
&&  \mathcal{I}^{\bm{e}}_\mu \equiv   \! \frac{1}{3} (D_\mu \theta  + \theta a_\mu)   \! + \!   \frac{3}{5} h^{\nu  \rho} \! \! \left(\! D_{( \rho} \sigma_{\mu \nu )}  +  a_{( \rho} \sigma_{\mu \nu )} \! \right) \! - \! 2 a^\nu \omega_{\mu \nu}  \, ,  \nonumber  \\   
&& \mathcal{I}^{\bm{d}}_\mu \equiv - 2  a_\mu \, ,   \nonumber  \\  
&& \mathcal{I}^{\bm{ee}}_{\mu \nu} \equiv - \left(4 \omega_{\alpha \mu}  \sigma^{\alpha}_{\; \nu} +  4 \omega_{\alpha \la \mu}  \omega^{\alpha}_{\; \nu \ra}    + D_{\la \nu} a_{\mu \ra} + a_{\la \mu} a_{\nu \ra } \right)   \, , \nonumber  \\   
&& \mathcal{I}^{\bm{ed}}_{\mu \nu} \equiv  4 (\sigma_{\mu \nu}  - \omega_{\mu \nu}   )  \, ,  \nonumber  \\   
&& \mathcal{I}^{\bm{eee}}_{\mu \nu \rho} \equiv   D_{\la \rho} \sigma_{\mu \nu \ra} +  a_{\la \rho} \sigma_{\mu \nu \ra}   \, , 
\eea  
where we have assumed that the null congruence is irrotational, such that $\nabla_{[\alpha} k_{\nu]} = 0$.  
The operator $D_\mu$ is the spatial covariant derivative\footnote{The acting of $D_\mu$ on a tensor field $T_{\nu_1 , \nu_2, .. , \nu_n }^{\qquad \quad \; \gamma_1 , \gamma_2, .. , \gamma_m }$ is defined as: $D_{\mu} T_{\nu_1 , \nu_2, .. , \nu_n }^{\qquad \quad \; \gamma_1 , \gamma_2, .. , \gamma_m } \equiv  h_{ \nu_1 }^{\, \alpha_1 } h_{ \nu_2 }^{\, \alpha_2 } .. h_{ \nu_n }^{\, \alpha_n }    \,  h_{ \beta_1 }^{\, \gamma_1 } h_{ \beta_2 }^{\, \gamma_2 } .. h_{ \beta_m }^{\, \gamma_m }    \, h_{ \mu }^{\, \sigma } \nabla_\sigma  T_{\alpha_1 , \alpha_2, .. , \alpha_n }^{\qquad \quad \; \beta_1 , \beta_2, .. , \beta_m }$ .
} 
on the 3-dimensional space orthorgonal to $\bm u$. 
The decomposition into multipoles in (\ref{integrandexp}) has been made such that the quadrupole moments $\mathcal{I}^{\bm{ee}}_{\mu \nu}$, $\mathcal{I}^{\bm{ed}}_{\mu \nu}$ and octupole moment $\mathcal{I}^{\bm{eee}}_{\mu \nu}$ are traceless. 
The multipole coefficients (\ref{coef}) are all given in terms of the kinematic variables of the time-like congruence and their first derivatives. 
The truncated expansion in (\ref{integrandexp}) is \emph{exact} and all higher order multipole contributions vanish by identity in the generic case.   
We note that $\mathcal{I}$ and $ \Sp_{\emi \rightarrow \obs}$ are invariant under rescalings of the affine parametrisation of the null congruence $\{\bm k \rightarrow \alpha \bm k , \lambda \rightarrow \lambda/\alpha \}$ with $k^\mu \nabla_\mu \alpha = 0$. This reflects the independence of the results on the emission lines of the astrophysical sources considered in the observational analysis. 

The set of equations (\ref{redshiftdriftdec}), (\ref{def:Eevolution}), (\ref{ederivative}), and (\ref{integrandexp}) constitute the main result of this paper. It provides a general expression for redshift drift in a physically interpretable multipole representation, where the multipole coefficients of the expansions (\ref{def:Eevolution}) of $\Eu$ and (\ref{integrandexp}) of $\mathcal{I}$ are given in terms of kinematic and dynamical variables of the observer congruence.

\section{Approximations in space-times with statistical symmetries}
In practice, cosmological information is deduced from the measurement of the redshift drift of many astrophysical emitters. 
When a notion of statistical homogeneity and isotropy is present and the fluid variables and their derivatives have no preferred orientation relative to the spatial direction of propagation $\bm e$ and its acceleration $\bm d$, we might expect monopole parts of $\mathcal{I}$ and $\Eu_\emi$ to be dominant in the collective analysis of the emitters\footnote{For cancellation of the higher order multipoles to occur, the length scales of light-propagation must be larger than the largest cosmological structures, and the astronomical sources considered must sample many uncorrelated cosmological structures. For further discussions on the cancellation of traceless contributions in statistical light propagation, see \cite{rae2009,rae2010}} such that the following approximation holds 
\bea
\label{redshiftdriftav}
\overline{\frac{d z}{d \tau_o} \Bigr\rvert_{\obs} } &=& (1+\overline{z}) \Eu_\obs  - \overline{\Eu_\emi} + \overline{\Sp_{\emi \rightarrow \obs}} \nonumber  \\   
&\approx& (1+\bar{z}) \Eu_\obs  - \frac{1}{3}\overline{\Phi_\emi}   \, , 
\eea  
where  
\bea
\label{thetaeff}
\overline{\Phi_\emi} \equiv \overline{\theta_\emi} - 3  \overline{E_\emi \! \!  \int_{\lambda_\emi}^{\lambda_\obs} \! \! \! d \lambda \, \mathcal{I}^{\it{o}}}    \, , 
\eea  
plays the role of an effective average expansion scalar, and where the overbar denotes averaging over objects contained within a suitable redshift interval. 
The contribution from the integral over the monopole $\mathcal{I}^{\it{o}}$ in (\ref{thetaeff}) can in general not be expected to be subdominant. In particular, the terms $-4\omega^{\mu \nu} \omega_{\mu \nu}/3$, $-a^{\mu} a_{\mu}/3$, and $-d^\mu d^\nu h_{\mu \nu}$ entering $\mathcal{I}^{\it{o}}$ are non-positive and generally result in accumulated negative contributions to the redshift drift\footnote{This result is consistent with the numerical findings of \cite{Koksbang:2020zej,Koksbang:2019glb} where the conjectured average approximation for redshift drift $(1+z) \overline{\theta_\obs}/3  - \overline{\theta_\emi}/3$ was considered for light propagation between ensembles of emitters and observers separated by the same distance in redshift. This approximation was found to yield too large values of the redshift drift as compared to the mean of the individually propagated light beams. Comparing with the expression (\ref{redshiftdriftav}), the overestimation of the average redshift drift in \cite{Koksbang:2020zej,Koksbang:2019glb} is likely due to ignoring the contributions from $\overline{\Sp_{\emi \rightarrow \obs}}$.} (or equivalently accumulated positive contributions to the effective average expansion scalar (\ref{thetaeff})). 
Redshift drift does thus in general {\em not} measure the (average) expansion rate $\theta$ of the congruence directly, also not in the case of statistically homogeneous and isotropic space-times: inhomogeneities and anisotropies on small scales enter the expression for redshift drift through $\Sp_{\emi \rightarrow \obs}$ with contributions that in general do not cancel on average.  
The constant offset $\Eu_\obs$ is determined by the coefficients $\{\theta/3, -a_\mu , \sigma_{\mu \nu} \} \rvert_{\obs}$ given by 9 degrees of freedom in total. For an observer situated in a volume with regional anisotropy, $a_\mu \rvert_{\obs}$ and $\sigma_{\mu \nu} \rvert_{\obs}$ can in general not be ignored. Dipolar and quadrupolar components in the measurements of average redshift drift are thus direct signatures of non-vanishing 4-acceleration and shear degrees of freedom respectively in the observers cosmological vicinity.

\section{Conclusion} 
\label{sec:conclusion} 
We have considered the general expression for redshift drift, as written in a physically interpretable multipole representation. 
The main result of this paper is given by the set of equations (\ref{redshiftdriftdec}), (\ref{def:Eevolution}), (\ref{ederivative}), and (\ref{integrandexp}) providing a general expression for redshift drift in terms of the truncated multipole series $\Eu$ and $\mathcal{I}$. The multipole coefficients of the series expansions are given in terms of the kinematic and dynamical variables of the observer congruence. The impact of regional inhomogeneity and anisotropy along the null rays for the measured redshift drift is transparent from the derived expression; in particular $\mathcal{I}$ contains monopole terms which are in general not expected to cancel along the light beam. 
An important conclusion to be drawn from the formulated expressions is that redshift drift is in general not a direct measurement of the (average) expansion rate -- also not in statistically homogeneous and isotropic space-times -- but depends on the configuration and dynamics of structure along the null rays. 
This illustrates the importance of accurately incorporating the existence of cosmological structures, even when light propagation over large distances is considered. This importance has also been discussed in the context of the luminosity distance--redshift relation \cite{Heinesen:2020bej}. 
In general the configuration of structure on small and intermediate scales enter effective large scale cosmological descriptions. This entanglement of scales is the core of the \sayy{fitting problem} of cosmology \cite{Ellis:1984bqf,Ellis:1987}. 

A prediction from the analysis presented in this paper, is that measurements of redshift drift in statistically homogeneous and isotropic space-times tend to acquire negative contributions from the integrated inhomogeneities along the lightrays.  
Another prediction is the presence of a dipole and quadrupole offset in the detected (average) redshift drift in space-times where the observer is located in a region of local anisotropy involving non-zero 4-acceleration and shear degrees of freedom. 

\vspace{6pt} 
%%%%%%%%%%%%%%%%%%%%%%%%%%%%%%%%%%%%%%%%%%%%%%%%%%%%%%%%%%%%%%%%%%%
\begin{acknowledgments}
%%%%%%%%%%%%%%%%%%%%%%%%%%%%%%%%%%%%%%%%%%%%%%%%%%%%%%%%%%%%%%%%%%%
This work is part of a project that has received funding from the European Research Council (ERC) under the European Union's Horizon 2020 research and innovation programme (grant agreement ERC advanced grant 740021--ARTHUS, PI: Thomas Buchert). I thank Thomas Buchert for his reading and comments on the manuscript. 
\end{acknowledgments}

%%%%%%%%%%%%%%%%%%%%%%%%%%%%%%%%%%%%%%%%%%%%%%%%%%%%%%%%%%%%%%%%%%%

\end{document}